\newcommand{\PaperTitle}{Watching TV with the Second-Party: A First Look at Automatic Content Recognition Tracking in Smart TVs}

\documentclass[sigconf]{acmart}

\copyrightyear{2024}
\acmYear{2024}
\setcopyright{rightsretained}
\acmConference[IMC '24]{Proceedings of the 2024 ACM Internet Measurement Conference}{November 4--6, 2024}{Madrid, Spain}
\acmBooktitle{Proceedings of the 2024 ACM Internet Measurement Conference (IMC '24), November 4--6, 2024, Madrid, Spain}
\acmDOI{10.1145/3646547.3689013}
\acmISBN{979-8-4007-0592-2/24/11}

\usepackage{xspace,soul}
\usepackage{subfigure}
\usepackage{url}
\usepackage{balance}

\newcommand{\etal}{et al.\xspace}

\begin{document}

\pagenumbering{gobble}

\title{\PaperTitle}


\author{Gianluca Anselmi}
\authornote{Both authors contributed equally to this research.}
\affiliation{
  \institution{University College London}
  \city{London}
  \country{United Kingdom}
}
\email{gianluca.anselmi.22@ucl.ac.uk}
\orcid{0009-0005-8500-9112}
\author{Yash Vekaria}
\authornotemark[1]
\affiliation{
  \institution{University of California, Davis}
  \city{Davis}
  \country{United States}
}
\email{yvekaria@ucdavis.edu}
\orcid{0000-0002-1891-7568}

\author{Alexander D'Souza}
\affiliation{%
  \institution{University of California, Davis}
  \city{Davis}
  \country{United States}
}
\email{aledsouza@ucdavis.edu}
\orcid{0009-0006-2464-9383}

\author{Patricia Callejo}
\affiliation{
  \institution{Universidad Carlos III de Madrid, uc3m-Santander Big Data Institute} 
  \city{Madrid}
  \country{Spain}
}
\email{pcallejo@it.uc3m.es}
\orcid{0000-0001-6124-6213}

\author{Anna Maria Mandalari}
\affiliation{
  \institution{University College London}
  \city{London}
  \country{United Kingdom}
}
\email{a.mandalari@ucl.ac.uk}
\orcid{0000-0002-6715-101X}

\author{Zubair Shafiq}
\affiliation{%
  \institution{University of California, Davis}
  \city{Davis}
  \country{United States}
}
\email{zubair@ucdavis.edu}
\orcid{0000-0002-4500-9354}

\renewcommand{\shortauthors}{Gianluca Anselmi and Yash Vekaria, \etal}

\begin{abstract}
Smart TVs implement a unique tracking approach called Automatic Content Recognition (ACR) to profile viewing activity of their users. 
ACR is a Shazam-like technology that works by periodically capturing the content displayed on a TV's screen and matching it against a content library to detect what content is being displayed at any given point in time.
While prior research has investigated third-party tracking in the smart TV ecosystem, it has not looked into \textit{second-party} ACR tracking that is directly conducted by the smart TV platform. 
In this work, we conduct a black-box audit of ACR network traffic between ACR clients on the smart TV and ACR servers. 
We use our auditing approach to systematically investigate whether (1) ACR tracking is agnostic to how a user watches TV (e.g., linear vs. streaming vs. HDMI), (2) privacy controls offered by smart TVs have an impact on ACR tracking, and (3) there are any differences in ACR tracking between the UK and the US. 
We perform a series of experiments on two major smart TV platforms: Samsung and LG. 
Our results show that ACR works even when the smart TV is used as a ``dumb'' external display, opting-out stops network traffic to ACR servers, and  there are differences in how ACR works across the UK and the US. 
\end{abstract}

\vspace{-2mm}

\ccsdesc[500]{Information systems~Content match advertising}
\ccsdesc[300]{Information systems~Online advertising}
\ccsdesc[300]{Information systems~Traffic analysis}
\ccsdesc[500]{Social and professional topics~Privacy policies}
\ccsdesc[300]{Social and professional topics~Corporate surveillance}
\ccsdesc[500]{Security and privacy~Privacy protections}
\ccsdesc[300]{Security and privacy~Social aspects of security and privacy}

\keywords{Smart TV, ACR, Fingerprinting, Advertising, Tracking, Privacy} 

\received{15 May 2024}
\received[revised]{5 September 2024}
\received[accepted]{11 September 2024}

\maketitle

\vspace{-3mm}
\section{Introduction}
\label{sec:intro}

Smart TVs, which can connect to the Internet and stream content, have become widely popular.
The smart TV penetration has reached almost a three-fourth of households today \cite{report_smart_tv, freewheelreport}, with a vast majority of globally sold TVs being smart~\cite{statistareport}.
In fact, it is challenging to buy a ``dumb'' TV now \cite{tomsguidereport}. 
A number of different smart TV platforms exist, led by Samsung Tizen and LG WebOS \cite{statistareport2}.

The research community has examined privacy issues in the smart TV ecosystem, particularly third-party tracking in smart TV apps~\cite{mohajeri2019watching, varmarken2020tv, tileria2022watch, tagliaro2023still}, but has not looked at second-party tracking conducted directly by the smart TV platform.
%
%
Smart TV platforms use a unique tracking approach dubbed Automatic Content Recognition (ACR) \cite{ACR} to profile viewing habits of smart TV users. 
Unlike traditional online tracking in the web and mobile ecosystems that is typically implemented by third-party libraries/SDKs included in websites/apps, ACR is typically directly integrated in the smart TV's operating system. 
At a high level, ACR works by periodically capturing the content displayed on a TV's screen and matching it against a content library to detect the content being viewed on the TV. 
It is essentially a Shazam-like technology for audio/video content on the smart TV \cite{themarkuparticle}.
%

ACR is implemented by all major smart TV manufacturers, including Samsung \cite{Samsung-ACR} and LG \cite{LG-ACR}. 
There has been public and regulatory scrutiny of ACR tracking.
Most notably, the FTC sued Vizio and Inscape in 2017 for their use of ACR tracking in smart TVs without user consent \cite{ftcsue}. 
However, prior research lacks an in-depth measurement and analysis of ACR tracking in smart TVs despite (a) being known to exist for many years, (b) its unique tracking approach as compared to web or mobile, and (c) its  deployment in the vast majority of smart TVs today.

Our research aims to bridge this gap in the prior literature. 
While it would be ideal to study ACR tracking in a white-box setting, it requires reverse-engineering and/or jailbreaking a smart TV's operating system, which is  challenging. 
An alternate auditing approach, commonly used in the measurement community \cite{iqbal2022your, moniotr_paper}, is to analyze network traffic flows between devices at the client-side and tracking endpoints at the server-side.
In this work, we adopt this auditing approach to analyze the network traffic between ACR clients on the smart TV and ACR servers for two major smart TV manufacturers: Samsung and LG.

We use this auditing approach to answer three research questions about ACR tracking in smart TVs.

\noindent \textit{\textbf{First}}, we investigate if ACR tracking is agnostic to how a user watches the TV. 
Specifically, we compare ACR tracking across scenarios where a user (a) watches linear TV (e.g., via antenna), (b) streams content from a smart TV platform's app such as Samsung TV Plus \cite{samsungtvplus}, (c) streams content from a third-party app such as Netflix, (d) uses the TV as an external display for laptop or gaming console via HDMI, (e) screen casts content via Wi-Fi from an external mobile phone or laptop, or (f) stays on the TV's homepage. 
ACR tracking across these scenarios raises unique concerns.  
For instance, ACR tracking when the TV is being used as a ``dumb'' external display raises privacy concerns. 
Similarly, ACR tracking of copyrighted third-party content raises intellectual property concerns \cite{CopyrightAct}.
We find that:
(1) ACR network traffic exists when watching linear TV and when using smart TV as an external display using HDMI, (2) ACR network traffic is not present when streaming content from third-party apps such as Netflix and YouTube.

\noindent \textit{\textbf{Second}}, we investigate whether privacy controls offered by smart TV manufacturers have an impact on ACR tracking. 
Prior research has shown that privacy controls do not always work \cite{liu2023opted, sanchez2019can, matte2020cookie}. 
We compare ACR tracking before and after exercising the offered privacy controls. 
We find that:
(1) opting-out stops ACR network traffic, (2) login status does not impact ACR network traffic.

\noindent \textit{\textbf{Third}}, we investigate whether ACR tracking differs across smart TVs bought and operated in different jurisdictions: the UK and the US. 
Both the UK and the US have distinct privacy laws and regulations. 
Moreover, data transfers between the UK and the US are regulated by the UK-US Data Bridge, which may impose geographic constraints on data transfers between smart TVs and ACR servers \cite{factsheetUK}. 
We study whether there are any differences in ACR tracking across the UK and the US and whether they use different geographically located ACR servers. 
We find that:
(1) smart TVs in the UK and the US contact distinct ACR domains, (2) unlike the UK, ACR is active in the US even when streaming content from the platform streaming app.

To the best of our knowledge, our work presents the first in-depth measurement and analysis of \textit{second-party} ACR tracking in smart TVs. 
For reproducibility, our code and data is available at: \url{https://github.com/SafeNetIoT/ACR}

\vspace{-3mm}
\section{Background and Motivation}
\label{sec:background}

Smart TV tracking can be broadly classified into second-party and third-party tracking. 
Third-party tracking on smart TV is similar to traditional web and mobile tracking.
Smart TV app developers include a tracking library or SDK that collects and shares app usage data and user/device identifiers with third parties.
In contrast, second-party tracking refers to the tracking conducted directly by the smart TV platform via its operating system. 
Unlike third-party tracking that is limited to the app that includes the tracking library, second-party tracking is potentially agnostic to how a user watches TV (i.e., whether watching linear or OTT, streaming app used, etc.).

\begin{figure}[t]
    \centering
    \includegraphics[width=\linewidth]{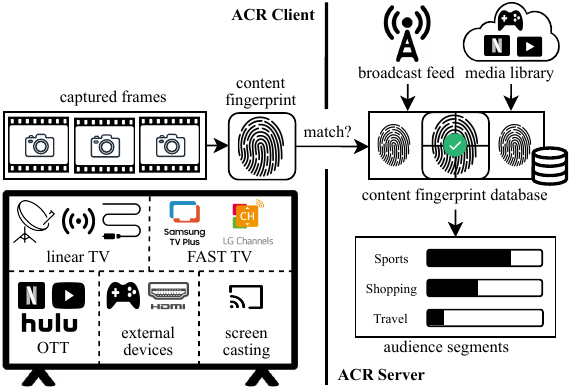} 
    \vspace{-4mm}
    \caption{Overview of ACR tracking in smart TVs.}
    \label{fig:acr}
    \vspace{-4mm}
\end{figure}

Automatic Content Recognition (ACR)~\cite{ACR} is widely used for second-party tracking in smart TVs. 
As shown in Figure \ref{fig:acr}, ACR periodically captures frames (and/or audio), builds a fingerprint of the content, and then shares it with an ACR server for matching it against a database of known content (e.g., movies, ads, live feed). 
When the fingerprint matches, ACR server can determine exactly what piece of content is being watched on the smart TV. 
This enables smart TV platforms like Samsung and LG to profile users into \textit{audience segments}~\cite{samsunghelpcenter,lgnewacrsolution}, which are then used to target personalized ads.
Fingerprints in ACR are essentially hash of the content, which can be matched at the server-side to identify the content. 
However, the fact that the hash of content rather than raw content is sent to ACR servers does not necessarily make the data ``anonymous''~\cite{ftc-hashing}. 
Moreover, the viewing habits of a user is potentially identifying \cite{narayanan2008robust}.

Since its inception in 2011, with roots in Shazam song identification, ACR tracking has been adapted to identify other modalities of content. 
In 2012, DirecTV and Viggle expanded ACR into the TV ecosystem~\cite{HistoryDirectTV}, while Samsung partnered with a content recognition tech company -- Enswers to integrate ACR into their smart TVs~\cite{HistoryEnswers}.
LG smart TVs incorporated ACR in 2013 with a partnership with Cognitive Networks~\cite{HistoryLGCognitiveNetworks}. 
The same year, Sony also partnered with Samba TV to use its own ACR~\cite{HistorySambaTV}. 
Moreover, Vizio and Roku adopted ACR in 2014~\cite{HistoryVizioTV}.

ACR tracking has raised privacy concerns. 
Most notably, Vizio was sued by the FTC for selling customer data to third parties, who then used it for personalized ads. 
This lawsuit was settled in 2017 with Vizio agreeing to provide clearer disclosures and opt-out mechanisms~\cite{HistoryVizioControversy}.
However, opting out is typically not straightforward, often requiring navigation through various settings in multiple subsections, with no universal off switch~\cite{TurnOffSmartTVFeatures}.
It is also unknown whether these privacy controls actually work as intended.

\section{Design \& Methodology}
\label{sec:system}

\subsection{Design}
\begin{figure}[t!]
	\centering
    \includegraphics[width=0.9\columnwidth]{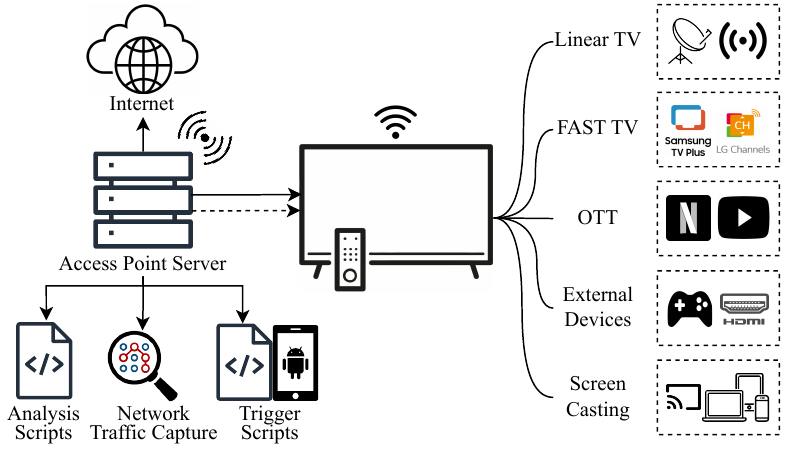} 
    \vspace{-0.5cm}
    \caption{Experimental setup.}
    \vspace{-0.5cm}
  \label{fig:system_design}
\end{figure}

To investigate ACR tracking, we setup a dedicated infrastructure that facilitates data collection and experimentation on smart TVs. 
Figure~\ref{fig:system_design} shows the design of our setup. 
We consider two smart TVs: Samsung and LG. 
We deploy our infrastructure in both the UK and the US.
Each component of our infrastructure is described below.

\noindent \textbf{Access Point Server}. The servers are the core of the infrastructure. Each server, one per TV, works as an access point for the smart TVs, using dedicated Wi-Fi cards or adapters. In addition, they have one wired network interface connected to the Internet, through which ACR domains are contacted.
The servers have been configured with the installation of the \textit{Mon(IoT)r} software~\cite{moniotr_paper, moniotr}, a powerful tool to capture network traffic from IoT devices.
These servers store all the captured traffic and the extracted data for further analysis.

\noindent \textbf{Smart TVs}. We select two smart TVs, specifically Samsung and LG (with 23\% and 18\% market share, respectively~\cite{tv-popularity}), making them two of the leading brands in the smart TV market, due to their widespread popularity and in-built integration of FAST channels.

\noindent \textbf{Scripts}. The experimentation process comprises a set of scripts, running directly on the servers, designed to automatically control the smart TVs while running different tests and analyzing their network traffic. 
In particular, we use support scripts for interacting with smart TVs and triggering a specific function, e.g., opening Netflix app (\textit{Trigger Scripts}) and verifying the correct execution of the experiments (\textit{Validation Scripts}). 
These scripts are entirely automated. 
\textit{Trigger Scripts} rely on Android mobile phones, physically connected to the servers.
We use the capabilities of Android Debug Bridge (ADB)~\cite{adb} to establish remote control over the Tuya Smart app~\cite{tuya_app}, effectively transforming mobile phones into remote controls for the smart TVs.
Finally, we use \textit{Analysis Script} for analysis of the network traffic.

\subsection{Methodology}
\noindent \textbf{Network Traffic Collection}.
Our servers are able to capture the encrypted traffic during the entire duration of our experiments. 
%
Our analysis is focused at extracting traffic patterns from the data captured by \textit{Mon(IoT)r} without decrypting it.
This tool allows us to design dedicated experiments tailored to each TV and scenario under investigation. 
The capture contains exclusively the traffic transmitted to and received from the smart TV.
The experimental workflow is as follows. 
After the initiation of traffic capture, we automatically power on the smart TV, using server-controlled smart plugs. 
This initial power-on phase is crucial, as the majority of DNS requests are typically sent within the first few seconds after device activation~\cite{thompson2021RapidIoTDevice}. 
This is essential to identify the domain names associated with the contacted IP addresses. 
After that, the core experiment starts, for a duration of one hour. 
The specific content of the experiment varies based on the scenario. 
Finally, the experiment concludes with the smart TV being powered off and the network traffic capture being terminated. 
The entire process is automated.

\noindent \textbf{Scenarios}. The term "scenario" refers to a distinct experimental setup designed to examine a specific functionality of the smart TV. 
Our experiments cover six scenarios.
\begin{itemize}
  
  \item \textbf{Idle}. The smart TV is powered on and remains on its home page for the entire duration of the experiment.
  
  \item \textbf{Linear}. The experiment involves watching a single linear channel broadcasted via the TV antenna.
  
  \item \textbf{FAST}. This involves streaming a single channel from the FAST platform of the TV manufacturer (Samsung TV+, LG Channels). 
  Free Ad-supported Streaming TV (FAST) is essentially linear broadcast TV that is streamed over the Internet. 
  
  \item \textbf{OTT}. An over-the-top (OTT) app, streaming app that provides streaming content over the Internet (Netflix or YouTube), is used to stream content.
  
  \item \textbf{HDMI}. A separate laptop (browsing and watching YouTube videos) or gaming console (playing popular games) is connected to the TV via HDMI.
  %
  
  \item \textbf{Screen Cast}. This scenario investigates the screen cast feature by mirroring YouTube content streamed on a separate phone or laptop onto the smart TV screen.

\end{itemize}

\begin{figure}[t!]
    \centering
    \includegraphics[width=0.86\columnwidth]{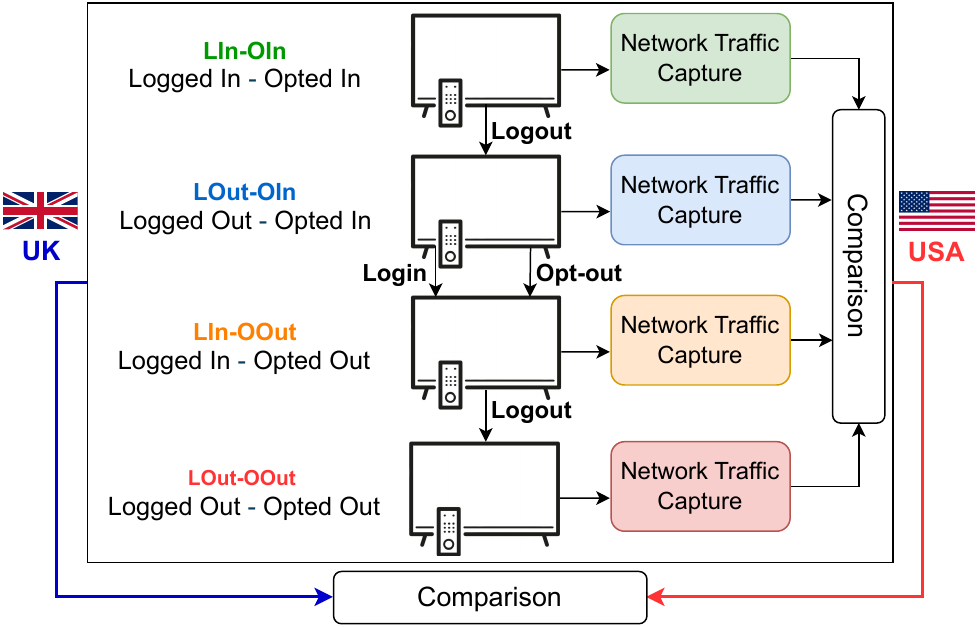}
     \vspace{-2mm}
     \caption{Methodology. The methodology is repeated for each scenario and TV in the two countries.}
     \label{fig:methodology}
     \vspace{-5mm}
\end{figure}

\noindent \textbf{Phases}. As shown in Figure \ref{fig:methodology}, we delineate four distinct phases for executing each scenario, each characterized by a unique configuration determined by two key factors: the linkage of a user account and the option to accept or reject advertising/tracking settings on the TV. 
In two of these phases, we are logged in using a TV account, while in the other two, we are logged out. Additionally, in two phases, we actively opt-out of all advertising/tracking options available directly on the TVs, thereby declining such services. Table \ref{tab:opt-out-options} in Appendix \ref{app:opt-out} lists all the selected opt-out options. Across all settings, ACR is specifically disabled by turning off \textit{viewing information services} \cite{tomsguidereport2}. Conversely, in the remaining two phases, we opt-in to such settings. 
It is important to note that without accepting the ToS and privacy policy we are unable to watch or access most of the Smart TV content. So, our opt-in or opt-out always assumes that ToS and privacy policy are opted-in.
Our four phases are as follows:

\begin{itemize}
  \item \textbf{LIn-OIn}. Logged In-Opted In
  \item \textbf{LOut-OIn}. Logged Out-Opted In
  \item \textbf{LIn-OOut}. Logged In-Opted Out
  \item \textbf{LOut-OOut}. Logged Out-Opted Out
\end{itemize}

\noindent \textbf{Identifying ACR traffic.} We employ the \textit{Analysis Scripts} to extract relevant information and statistics from the captured network traffic. 
%
%
We primarily focus on domains potentially associated with ACR tracking. 
To this end, we filter the list of contacted domains, identified with DNS captured packets, retaining only those containing the string "acr". 
%
To the best of our knowledge, no official documentation exists on the specific domain names used for ACR by Samsung or LG, nor is there a standard requiring all domains with "acr" in the name to be related to ACR on smart TVs. But we use this approach to identify ACR traffic for the following reasons:
\begin{itemize}
    \item Identified domains with the "acr" string were classified as tracking-related by sources like Netify~\cite{netify} and Blocada~\cite{blocada}. Blocada.org~\cite{blocada}, an open-source privacy suite, lists all such domains for LG and Samsung as tracking-related, matching the patterns we observed~\cite{blocada}.

    \item Numbered ACR domains we observed suggest a consistent naming scheme by LG and Samsung, likely used to distinguish ACR servers by region or other factors.

    \item We further validated our approach by comparing presence of these domains before and after opting-out of ACR on the Smart TV and analyzing contact frequency across scenarios. These domains showed regular contact patterns, unlike other ad/tracking domains like samsungads.com.
\end{itemize}

This approach however, may not generalise to all TV manufacturers. Future research can build on our findings by following the heuristic in the last point. Our automated methodology and released scripts support further exploration in this area.
\section{Results}
\label{sec:results}

The findings within the first two subsections pertain exclusively to the UK. 
The final subsection compares these results against the US. 

\vspace{-2mm}
\subsection{Comparison Across Scenarios}

To understand ACR tracking, we compare the network traffic flows to ACR domains during different scenarios, focusing particularly on phase LIn-OIn. 
LIn-OIn likely represents the most common configuration in the wild, since smart TV users need to be logged in (e.g., LG TVs require login for app downloads) and opted in (default option when setting up the TV) to access most of its functionalities. 

Our analysis reveals different behaviors between Samsung and LG regarding their use of ACR domains. 
When ACR is enabled on LG TVs, a single domain is contacted (\url{eu-acrX.alphonso.tv}, where X is an arbitrary number that changes periodically). 
This domain belongs to Alphonso, a technology company that manages LG Ad Solutions \cite{alphonso-terms-of-service}. 
%
%
On the other hand, Samsung contacts multiple ACR domains (\url{acr-eu-prd.samsungcloud.tv}, \url{acr0.samsungcloudsolution.com}, \url{log-config.samsungacr.com}, \url{log-ingestion-eu.samsungacr.com}).
All these domains belong to Samsung, aligning with the fact that Samsung Ads offer its own proprietary ACR tracking \cite{samsung-ads-tos}.
Furthermore, we geolocate the IP addresses using MaxMind \cite{maxmind} and IP2Location \cite{ip2location}.
%
Due to known limitations and inaccuracies of GeoIP databases, we perform additional validation using RIPE IPmap~\cite{ripe-ipmap} to accurately map the observed ACR domains to their server locations. 
We first perform traceroute from a location in the US or UK, then use RIPE IPmap for geolocation. 
In case of discrepancies, we rely on RIPE IPmap because -- 
(1) It offers multiple geolocation engines, each with unique techniques. 
(2) Its latency engine quickly computes measurements using RIPE Atlas probes with known locations. 
(3) It uses a reverse DNS engine that leverages geographical identifiers in PTR records to estimate IP locations.
%
%
Our analysis reveals that all LG ACR domains resolve to Amsterdam, Netherlands. 
Network traffic between the UK and the EU raises no cross-jurisdictional regulatory concerns~\cite{flowsukeu}.
For Samsung, \url{acr-eu-prd.samsungcloud.tv} and \url{log-ingestion-eu.samsungacr.com} are both located in London, UK, while \url{acrX.samsungcloudsolution.com} locates to Amsterdam, Netherlands, and \url{log-config.samsungacr.com} in New York, USA. 
This raises concerns about UK TV users' viewership data being stored in the US, where different privacy regulations apply. 
However, both Alphonso (for LG) and Samsung are on the DPF (Data Privacy Framework) List~\cite{dfplist, factsheetUK}, allowing data transfers between the UK and the US under the UK-US Data Bridge.

\begin{figure*}[h]
	\centering
    \subfigure[LG]{
        \includegraphics[width=0.9\columnwidth]{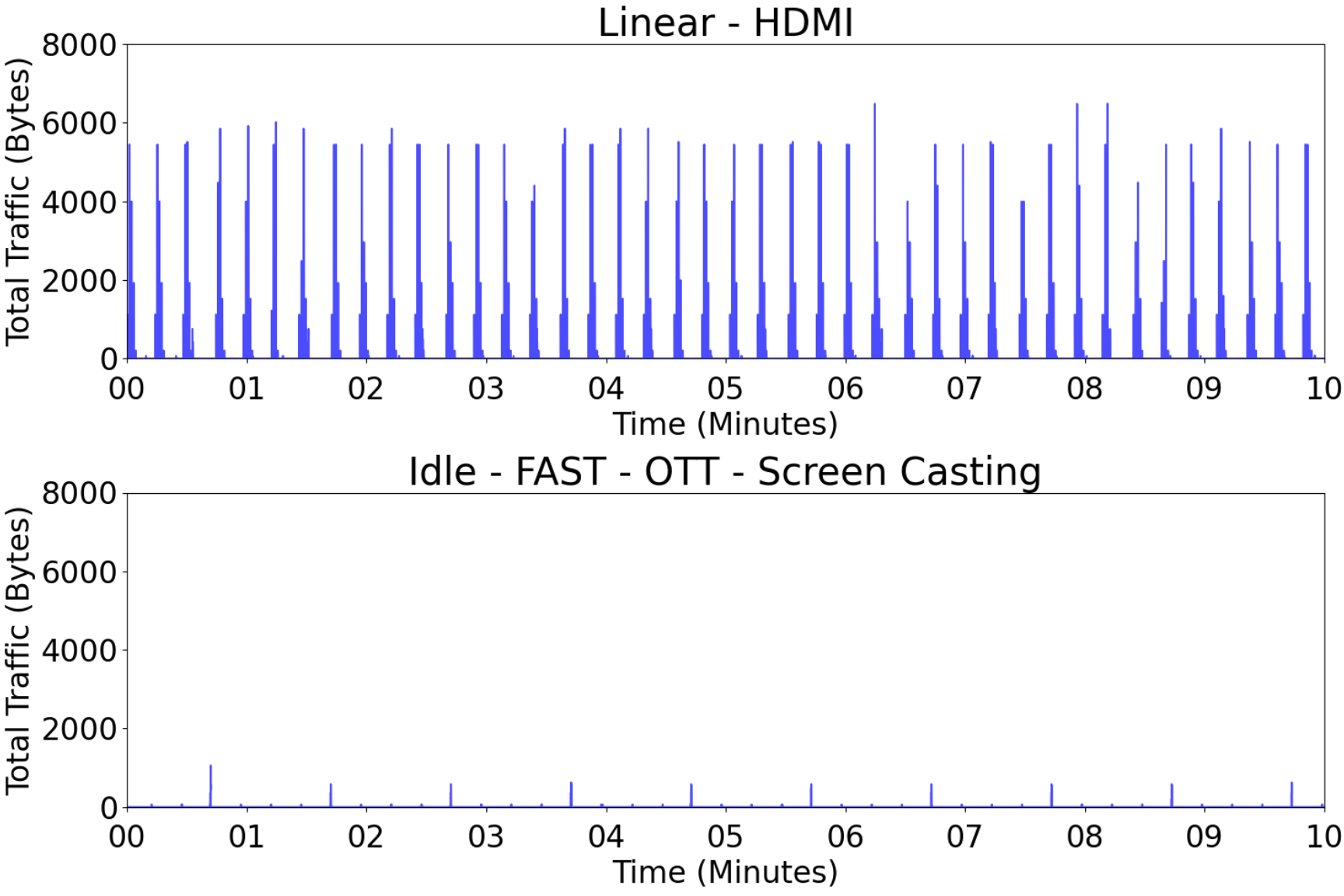}  
    }
	\centering
    \subfigure[Samsung]{
        \includegraphics[width=0.9\columnwidth]{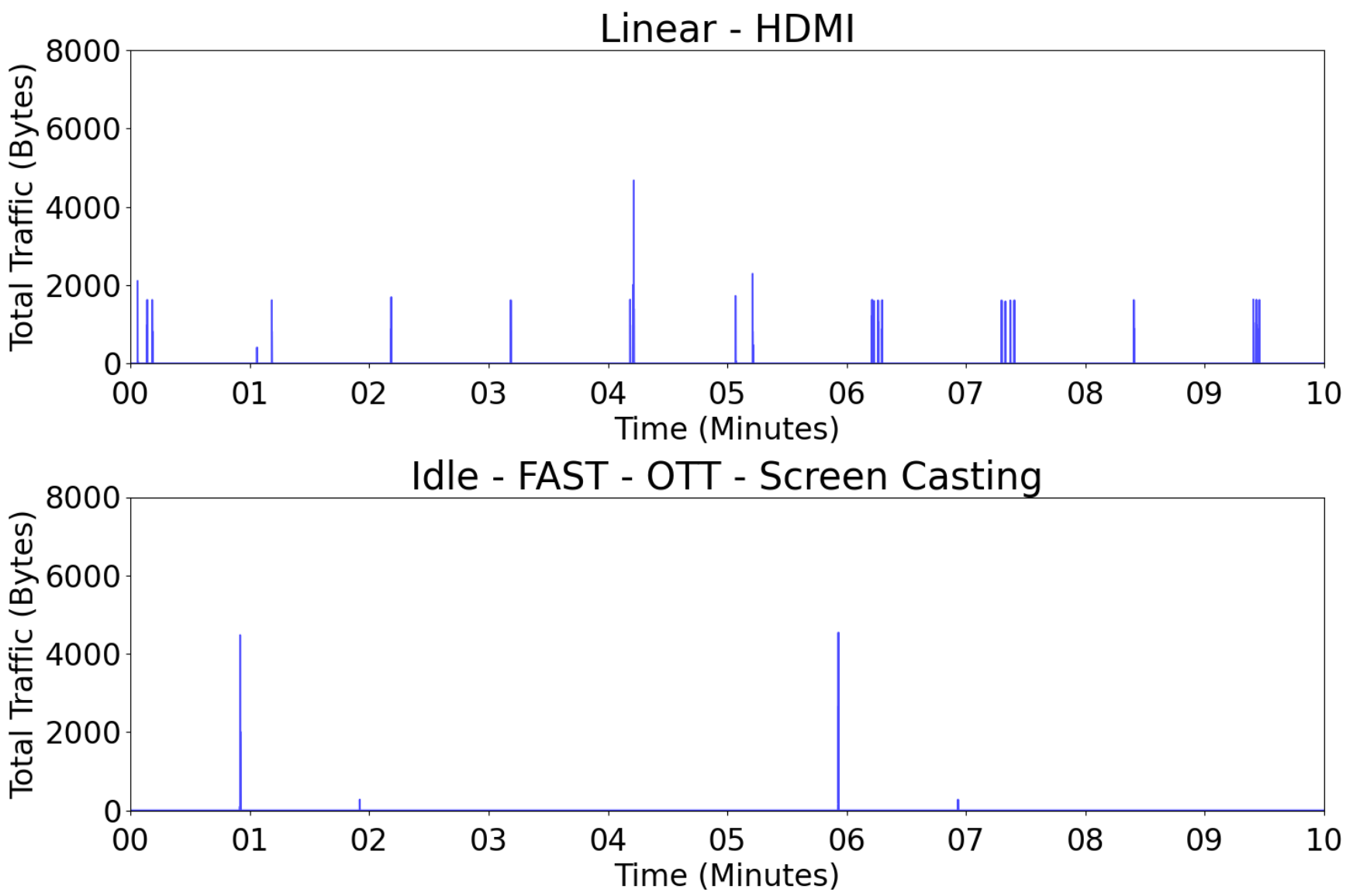} 
    }
    \vspace{-0.5cm}
    \caption{10 minutes of ACR traffic in different scenarios during LIn-OIn in UK.}
    \vspace{-0.4cm}
    \label{fig:P1}
\end{figure*}

Figure~\ref{fig:P1} shows the frequency of network traffic directed towards ACR domains in all scenarios. The data is presented in a packet-per-millisecond format, where each spike corresponds to a single millisecond slot. 
For scenarios in which similar behaviors are observed, only one plot is presented. 
%
For both LG (a) and Samsung (b) TVs, the scenarios with the highest ACR traffic are Linear and HDMI. 
%
%
During the remaining scenarios, ACR traffic is significantly less -- peaks get reduced by up to 12$\times$ -- suggesting that ACR client on the TV may not be sending fingerprints. 
%
%
%
%
LG's official documentation mentions that its ACR captures frames every 10ms~\cite{lg-privacy-policy}.
However, the fact that we observe network traffic every 15 seconds suggests that LG likely batches the frames captured every 10ms and sends the resultant content fingerprint every 15 seconds.
The remaining scenarios for LG show a lower amplitude of communication  occurring every fifteen seconds, with peaks every minute. 
%
%
For Samsung, we observe a consistent amount of traffic across various scenarios for the ACR domains (\url{acr0.samsungcloudsolution.com}, \url{log-config.samsungacr.com}, \url{log-ingestion-eu.samsungacr.com}). 
Based on their names, we hypothesize that communication with these domains primarily consists of logging information or maintaining connection persistence (``keep-alives"). 
We assume that the domain \url{acr-eu-prd.samsungcloud.tv} transmits fingerprints, as it exhibits the highest network traffic during Linear and HDMI. 
Communication occurs once per minute, with peaks observed approximately every five minutes. Interestingly, the remaining scenarios exhibit consistent peak values occurring every minute, accompanied by additional smaller traffic one minute following each peak.
%
%
Samsung's official documentation \cite{sm-privacy-policy} mentions that its ACR captures the frames every 500ms, suggesting that Samsung batches the captures as well and sends the fingerprints every minute. 
The differences in ACR capture frequency explains the different network behavior across the two brands.

%
We assume that, during OTT, ACR may not collect screenshots of third-party owned streamed content due to copyright issues.
Another explanation could be that the third-party app wants to preserve the privacy of its users, for example Netflix prefers to have ACR deactivated during its streaming ``in order to preserve the integrity of its subscribers viewing experiences and maintain sole control over measurement of its viewership" \cite{netflix-acr}.
The same reasoning applies to FAST channels, which LG and Samsung consider to be "aggregator apps"~\cite{aggregator-apps}, where providers may have agreements restricting ACR usage.
%
%
%

Linear and HDMI do not seem to present the previous restrictions. 
Linear, as FAST, may feature content from multiple providers, but the broadcasting network typically holds the rights to the content and has agreements in place with content owners and advertisers. 
These agreements include provisions for the use of ACR.

\vspace{-2mm}
\subsection{Comparison Across Phases}

This section describes the methods we use for understanding impact of privacy controls, specifically in terms of the influence of user login status and opt-out settings on ACR network traffic. 
%

\begin{figure*}[h]
    \centering
    \includegraphics[width=0.8\linewidth]{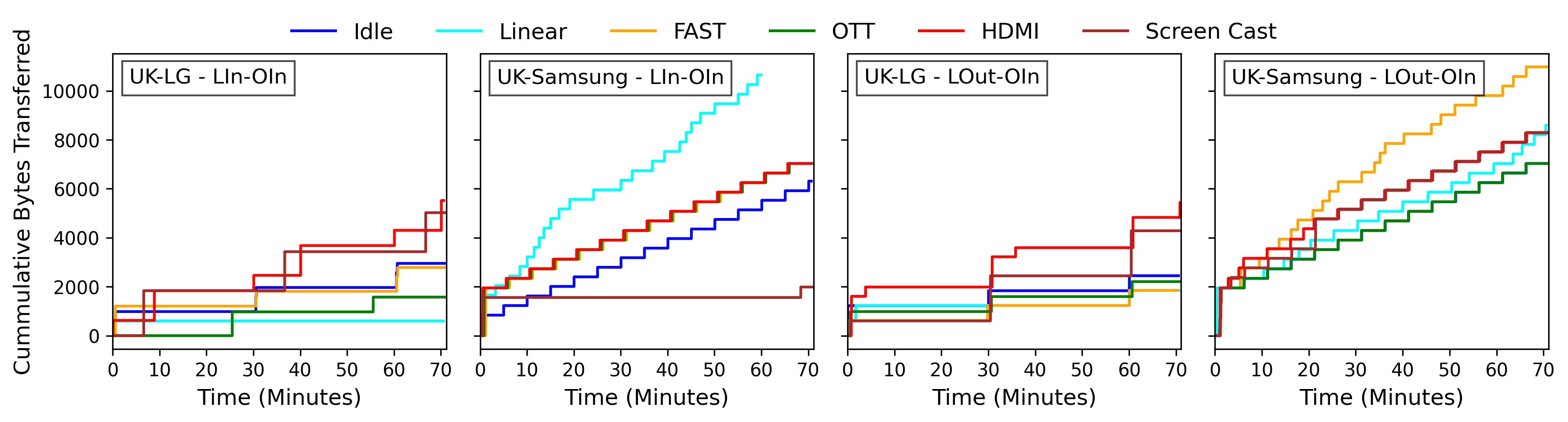}
    \vspace{-0.4cm}
    \caption{Cumulative distribution of bytes transmitted to ACR domains over the time during different opted-in phases in UK.}
    \vspace{-0.3cm}
    \label{fig:uk_bytes_transferred}
\end{figure*}

LIn-OIn differs from LOut-OIn by only that the user is logged in/out on the TV. 
%
We first understand the differences during these phases across TVs by plotting the CDF of data transferred to ACR domains (in bytes) in each scenario during the LIn-OIn and LOut-OIn phases as shown in Figure~\ref{fig:uk_bytes_transferred} (see Figure~\ref{fig:us_bytes_transferred} for USA). 
We observe distinctions in the data transfer periodicity amongst LG and Samsung measurements as also captured in Figure~\ref{fig:P1}. 
Overall, Samsung transmits upto $2X$ more data at a higher frequency to ACR domains as compared to LG. 
Interestingly, LG sends the most data to ACR endpoints when content is streamed via HDMI and screen casting, while Samsung does so during linear TV streaming in the logged-in phase and FAST in the logged-out phase.
These findings suggest implementational differences in the content fingerprint generation and transmission algorithms used by the two TVs.

Next, we look at the differences between the logged-in and the logged-out status for the same TV manufacturer. Analysis reveals that the set of ACR domains contacted across scenarios in LOut-OIn remains identical to those observed in LIn-OIn. 
Traffic volume and frequency patterns also exhibit a high degree of similarity between these phases for the same TV manufacturer as also represented in Figure~\ref{fig:uk_bytes_transferred}. 
%
%
Hence, we conclude that although differences exist across TV manufacturers, for a given TV, user login status appears to have no material impact on the ACR network traffic behavior. 
We also assume that ACR tracking may be relying on the Advertising ID of the TV and/or the IP address rather than the user account ID. 
For completeness, we add more details in Tables \ref{tab:bytes_acr} and \ref{tab:bytes_acr_P2} in Appendix \ref{app:bytes}.

Similarly, LIn-OOut and LOut-OOut phases exhibit identical behavior. 
They differ from the previous two in that we have opted out of all advertising/tracking. 
Interestingly, once opt-out is exercised (Table \ref{tab:opt-out-options}), there is a complete absence of communication with any previously identified ACR domains, and no new ACR-related domains are observed. 
These findings suggest that the opt-out mechanisms implemented on LG and Samsung smart TVs are working.

\begin{figure*}[!t]
	\centering
    \subfigure[LG]{
        \includegraphics[width=0.9\columnwidth]{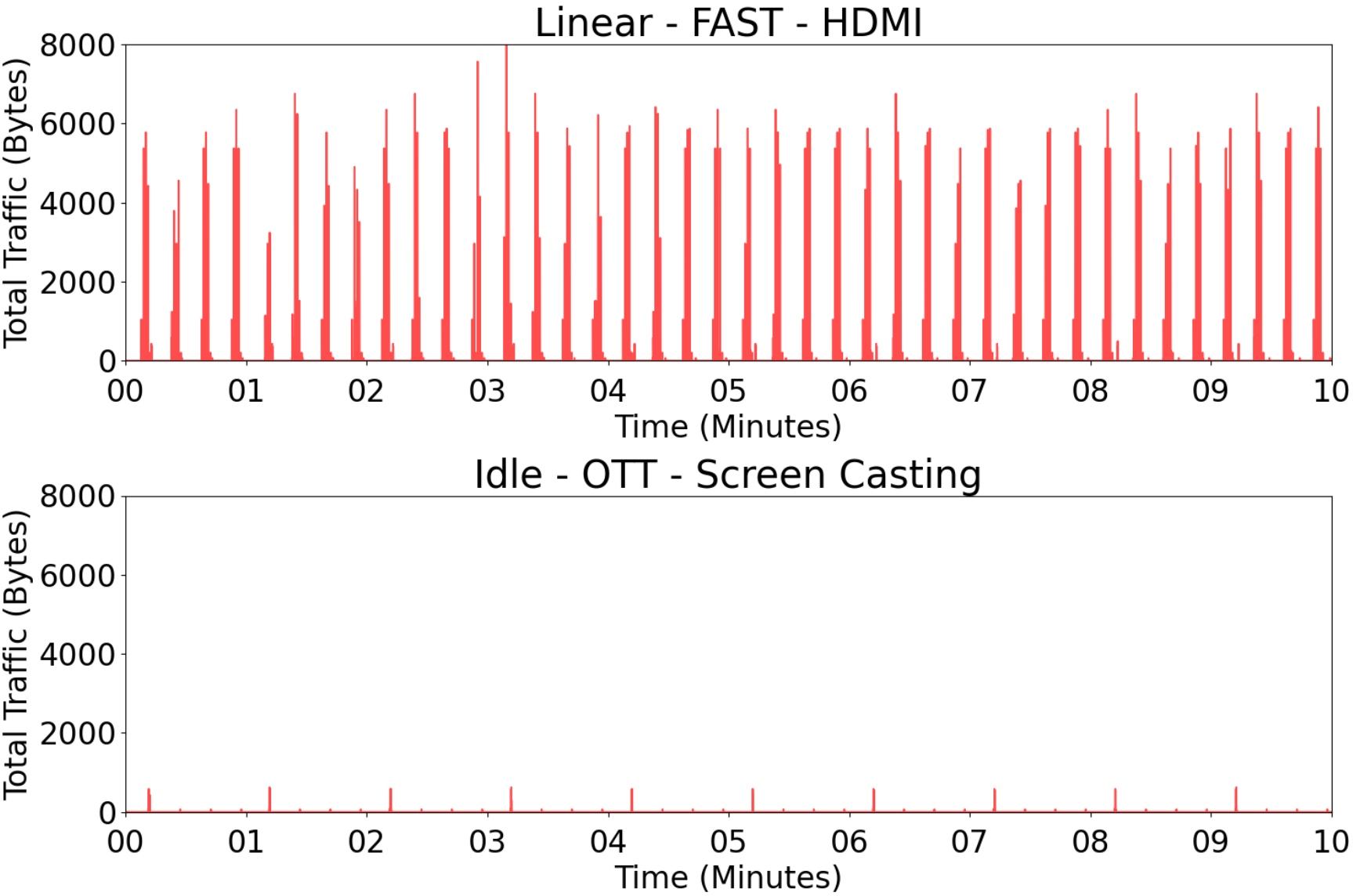}  
    }
	\centering
    \subfigure[Samsung]{
        \includegraphics[width=0.9\columnwidth]{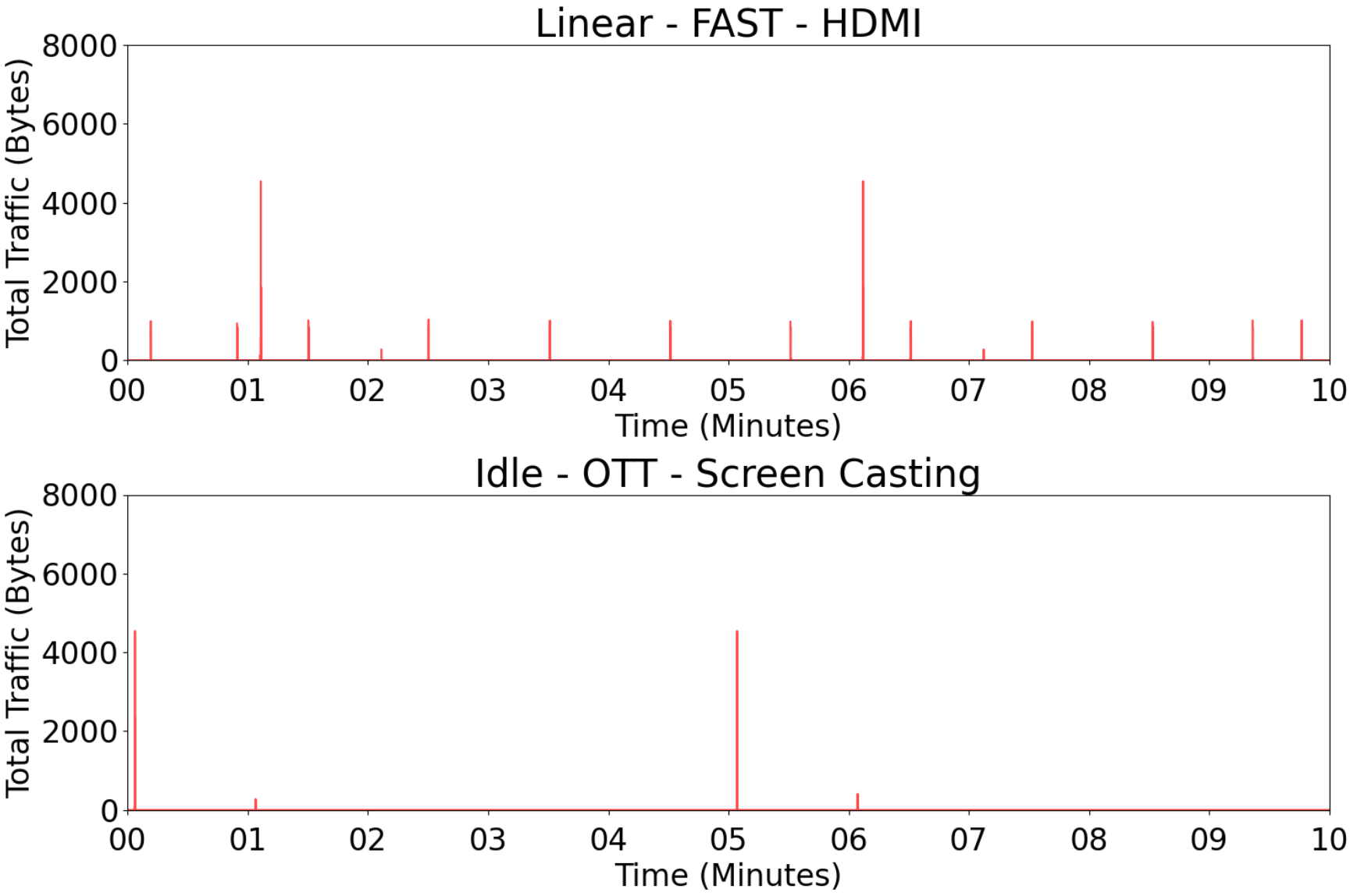} 
    }
    \vspace{-0.5cm}
    \caption{10 minutes of ACR traffic in different scenarios during LIn-OIn in US.}
    \vspace{-0.2cm}
    \label{fig:P1US}
\end{figure*}

\vspace{-3mm}
\subsection{Comparison Across Countries}

To investigate the differences across the US and the UK, we now analyze the results of our experiments conducted in the US. 
The comparison reveals some key differences in the ACR tracking across the US vs. the UK.
This comparison is particularly relevant due to the differing data protection laws: the GDPR (General Data Protection Regulation) in the UK~\cite{gdpr} and CCPA (California Consumer Privacy Act) in the US~\cite{ccpa}, particularly in California where we conducted out US experiments. 
It's possible that these variations in data protection laws may influence how ACR tracking operates in each region.
More details on network traffic to US-specific ACR domains are provided in Table \ref{tab:bytes_acr_US_P1} in Appendix \ref{app:bytes}.

Both LG and Samsung TVs utilize domain names consistent with those identified in the UK, with some differences in the names (e.g. the term ``EU'' in the UK domains and ``US'' in the US domains related to ACR). 
LG contacts a single ACR domain, \url{tkacr}\textbf{X}\url{.alphonso.tv} (where \textbf{X} is a number that changes periodically). 
Samsung, however, contacts the same three domains (\url{acr-us-prd.samsungcloud.tv}, \url{log-config.samsungacr.com}, and \url{log-ingestion.samsungacr.com}), but omits the fourth domain that it uses in the UK. 
Analyzing the geolocation of the US domains, their IP addresses all belong to servers that are physically located in the US. 
%

As shown in Figure~\ref{fig:P1US}, in the US -- similar to the UK -- both smart TV platforms exhibit a significantly higher network traffic with ACR domains during Linear, FAST, and HDMI scenarios. 
Idle, OTT, and Screen Casting display considerably less traffic with ACR domains. 
Interestingly, the FAST scenario deviates from the UK findings. 
Watching LG Channels and Samsung TV+ in the US results in ACR traffic levels comparable to linear channel viewing, possibly because FAST platforms have different agreements with the content providers than the ones in UK, allowing ACR to operate. 
All other assumptions made for the UK are valid also for the US, due to the similarity of the behaviors. 

For all four phases, the observations are identical in both countries.
In both LIn-OIn and LOut-OIn phases, user login status seems to have no influence on ACR tracking.
All traffic patterns for the first two phases in both the countries during all scenarios are shown in Figures \ref{fig:UKP1-all}, \ref{fig:UKP2-all}, \ref{fig:USP1-all} and \ref{fig:USP2-all} in Appendix \ref{app:bytes}, for completeness.
Conversely, opting out of advertisement/tracking (LIn-OOut and LOut-OOut) disables ACR in both the countries. 
There is a complete absence of communication with any ACR domains during these phases.

\vspace{-0.1cm}
\section{Related Work}
\label{sec:related}

Online advertising and tracking pervades web, mobile, and IoT ecosystems. 
Research community has investigated advertising and tracking in IoT devices~\cite{aksu2018advertising} emphasizing on either the IoT traffic~\cite{mazhar2020characterizing,huang2020iot,bello2017network} or security and privacy implications of such tracking~\cite{jin2022exploring,barbosa2019if,fawaz2019security,zheng2018user,pishva2017internet,iqbal2022your,becker2019tracking,trimananda2022ovrseen}.
%

Advertising and tracking in  smart TVs by third-parties has also garnered attention from the research community. 
%
%
%
In the last few years, researchers have  attempted to study advertising and tracking in the smart TV ecosystem~\cite{varmarken2020tv,mohajeri2019watching,tileria2022watch,varmarken2022fingerprintv,tagliaro2023still,gilbert2019there,le2021characterizing,moghaddam2022tracking}. 
Varmarken \etal~\cite{varmarken2020tv} developed a tool to collect and analyze network traffic from top-1000 apps on smart TVs -- Roku and Amazon Fire TV. 
They also show ineffectiveness of DNS-blocklists in blocking advertising and tracking traffic and PII-exposure by the ecosystem. 
Mohajeri \etal~\cite{mohajeri2019watching} studied the same two OTT streaming devices by analyzing their network traffic. 
They found that tracking involved collection and sharing of unique identifiers such as device IDs, serial numbers, MAC addresses, and SSIDs. 
Tileria \etal~\cite{tileria2022watch} showed the existence of similar tracking ecosystem for Android TVs. 
Tagliaro \etal~\cite{tagliaro2023still} studied security and privacy issues for the Hybrid Broadcast Broadband TV (HbbTV) standard that allows broadcasters to improve their offered content to the broadcast signal as well as  OTT streaming app users. 
While smart TV third-party tracking has been extensively studied, second-party tracking (e.g., ACR) has received little attention. 
We fill the gap in the literature by analyzing ACR technology in smart TVs.


\vspace{-2mm}
\section{Conclusion}
\label{sec:conclusion}

We present a first look at second-party Automatic Content Recognition (ACR) tracking in smart TVs.
Using a black-box auditing approach, we tested two major smart TV brands (LG and Samsung) in various scenarios and experimental setup in two different countries (UK and US).
Our findings indicate that (1) ACR operates even when it is used as a ``dumb'' display via HDMI; (2) opt-out mechanisms stop ACR traffic; (3) ACR works differently in the UK as compared to the US.
As future work, we plan to explore more advanced man-in-the-middle (MITM) techniques to understand the payload of ACR network traffic. 
Moreover, we plan to investigate the link between ACR tracking and ad personalization in smart TVs.
Finally, although different than ACR, our auditing approach can be adopted to assess privacy risks of Recall~\cite{recall} -- which analyzes snapshots of the screen using generative AI~\cite{recall-verge}.
To foster future research, our code and data is available at \url{https://github.com/SafeNetIoT/ACR}.

%

\vspace{-2mm}
\begin{acks}
This work is supported in part by Engineering and Physical Sciences Research Council award EP/S035362/1, the project AUDINT (Grant TED2021- 132076B-I00) funded by the MCIN/AEI/10.13039/\\501100011033 and the EU FEDER funds, and National Science Foundation awards CNS-2103038, CNS-2138139, and CNS-2103439.  
\end{acks}

\bibliographystyle{ACM-Reference-Format}
\balance
\bibliography{refs}

\appendix
\begin{table}[h]
    \small 
    \begin{tabular}{p{1cm} p{6.8cm}}
       \toprule
       \textbf{SmartTV} & \textbf{Opt-Out Option} \\[1mm] 
       \toprule
       \centering LG 
       & \textbf{Enable} \textit{Limit ad tracking} \\
       & \textbf{Disable} \textit{TV membership agreement for marketing comms.} \\
       & \textbf{Enable} \textit{Do not sell my personal information} \\
       & \textbf{Edit} \textit{User agreements} in Privacy and Terms as follows: \\ 
       & \hspace{2mm} \textbf{Disable}: \\
            & \hspace{2mm} \textit{-- Viewing information agreement}\\
            & \hspace{2mm} \textit{-- Voice information agreement}\\
            & \hspace{2mm} \textit{-- Interest-based \& Cross-device advertising agreement}\\
            & \hspace{2mm} \textit{-- Who.Where.What?} \\
       & \textbf{Disable} \textit{Home promotion} \\
       & \textbf{Disable} \textit{Content recommendation} \\
       & \textbf{Disable} \textit{Live plus} \\
       & \textbf{Disable} \textit{AI recommendation} (\textit{Who.Where.What}, \textit{Smart Tips}) \\[1mm] 
       \midrule
       \centering Samsung 
       & \textbf{Disable} \textit{I consent to viewing information services on this device} \\
       & \textbf{Disable} \textit{I consent to interest-Based advertisements} \\
       & \textbf{Disable} \textit{Customization Service} \\
       & \textbf{Enable} \textit{Do not track} \\
       & \textbf{Disable} \textit{Improve personalized ads} \\
       & \textbf{Disable} \textit{Get news and special offer} \\
       \bottomrule
    \end{tabular} \\[1mm]
    \caption{Opt-Out Options in the Smart TVs.}
    \label{tab:opt-out-options}
\end{table}

\begin{figure*}[h]
    \centering
    \includegraphics[width=0.85\linewidth]{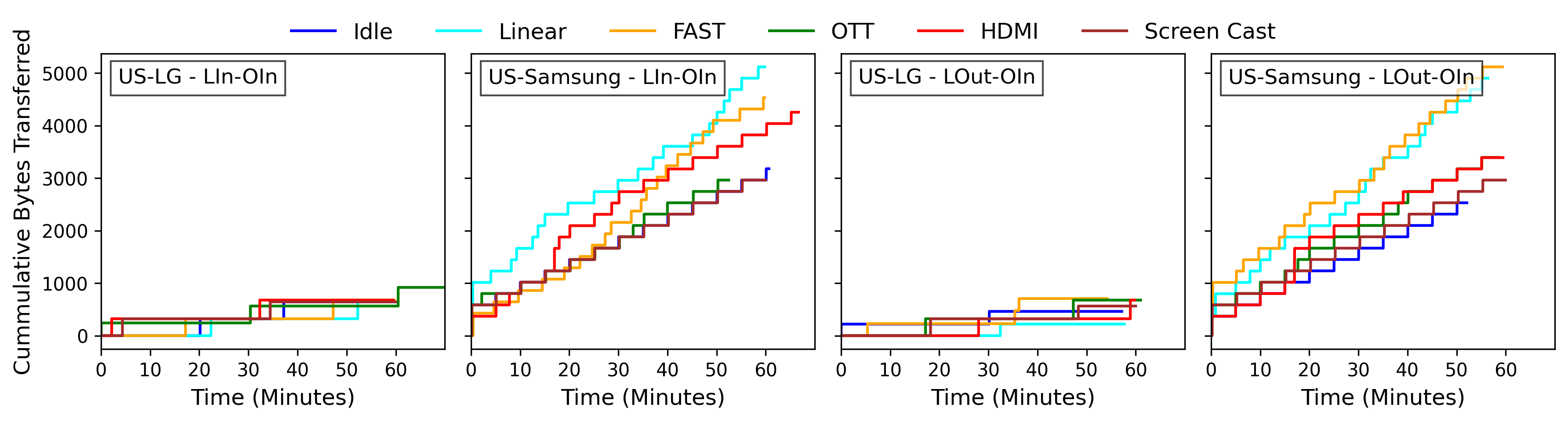}
    \caption{Cumulative distribution of bytes transmitted to ACR domains over the time during different opted-in phases in USA.\\}
    \label{fig:us_bytes_transferred}
\end{figure*}

\begin{table*}[h]
    \centering
    \begin{tabular}{cccccccc}
       \textbf{Domain Name} & \textbf{Idle} & \textbf{Antenna} & \textbf{FAST} & \textbf{OTT} & \textbf{HDMI} & \textbf{Screen Cast} \\ \hline
       \url{eu-acrX.alphonso.tv} & 264.7 & 4759.7 & 262.8 & 264.3 & 4296.5 & 266.2 \\ \hline
       \url{acr-eu-prd.samsungcloud.tv}  & - & 440.9 & 8.5 & 8.6 & 204.8 & 30.3 \\
       \url{acr0.samsungcloudsolution.com}  & - & - & 11.1 & 11.3 & 11.0 & 11.7 \\
       \url{log-config.samsungacr.com}  & 9.5 & 10.8 & 9.2 & 8.9 & 9.3 & 10.0 \\
       \url{log-ingestion-eu.samsungacr.com}  & 176.9 & 298.4 & 125.4 & 161.6 & 162.3 & - \\
    \end{tabular}
    \caption{Number of kilobytes sent/received to/from ACR domains in different scenarios during LIn-OIn in UK.}
    \label{tab:bytes_acr}
\end{table*}

\begin{table*}[h]
    \centering
    \begin{tabular}{cccccccc}
       \textbf{Domain Name} & \textbf{Idle} & \textbf{Antenna} & \textbf{FAST} & \textbf{OTT} & \textbf{HDMI} & \textbf{Screen Cast} \\ \hline
       \url{eu-acrX.alphonso.tv} & 258.0 & 4801.9 & 255.5 & 250.6 & 4229.5 & 272.8 \\ \hline
       \url{acr-eu-prd.samsungcloud.tv}  & 8.6 & 463.9 & 8.6 & 8.5 & 184.0 & 16.1 \\
       \url{acr0.samsungcloudsolution.com}  & 11.1 & 11.1 & 11.0 & 11.1 & 11.0 & 24.3 \\
       \url{log-config.samsungacr.com}  & 9.2 & 9.1 & - & 9.1 & 9.2 & 10.4 \\
       \url{log-ingestion-eu.samsungacr.com}  & 159.9 & 232.3 & - & 169.8 & 170.6 & 195.3 \\
    \end{tabular}
    \caption{Number of kilobytes sent/received to/from ACR domains in different scenarios during LOut-OIn in UK.}
    \label{tab:bytes_acr_P2}
\end{table*}

\begin{table*}[h]
    \centering
    \begin{tabular}{ccccccccc}
       \textbf{Domain Name} & \textbf{Idle} & \textbf{Antenna} & \textbf{FAST} & \textbf{OTT} & \textbf{HDMI} & \textbf{Screen Cast} \\ \hline
       \url{tkacrX.alphonso.tv} & 215.3 & 4583.2 & 4948.3 & 214.9 & 4125.0 & 240.4  \\ \hline
       \url{acr-us-prd.samsungcloud.tv}  & - & 184.4 & 176.6 & - & 148.5 & - \\
       \url{log-config.samsungacr.com}  & 10.5 & 10.5 & - & 9.7 & 19.7 & 10.1 \\
       \url{log-ingestion.samsungacr.com}  & 143.5 & 253.2 & 237.4 & 156.1 & 224.8 & 172.1 \\
    \end{tabular}
    \caption{Number of kilobytes sent/received to/from ACR domains in different scenarios during LIn-OIn in US}
    \label{tab:bytes_acr_US_P1}
\end{table*}

\begin{table*}[h]
    \centering
    \begin{tabular}{ccccccccc}
       \textbf{Domain Name} & \textbf{Idle} & \textbf{Antenna} & \textbf{FAST} & \textbf{OTT} & \textbf{HDMI} & \textbf{Screen Cast} \\ \hline
       \url{tkacrX.alphonso.tv} & 236.3 & 4612.4 & 4832.5 & 191.3 & 4633.5 & 222.0 \\ \hline
       \url{acr-us-prd.samsungcloud.tv}  & - & 153.5 & 166.1 & - & 160.2 & - \\
       \url{log-config.samsungacr.com}  & 9.6 & 9.6 & 9.6 & 10.4 & 10.4 & 9.6 \\
       \url{log-ingestion.samsungacr.com}  & 112.7 & 216.3 & 247.5 & 187.5 & 146.9 & 157.9 \\
    \end{tabular}
    \caption{Number of kilobytes sent/received to/from ACR domains in different scenarios during LOut-OIn in US}
    \label{tab:bytes_acr_US_P2}
\end{table*}

\section{Ethics}
 In our experiments we do not collect any data from real users on the Internet. All experiments are contained within our own testbed. When conducting the experiments, we fully respected the ethical guidelines defined by our affiliated organization, and we received approval.

\section{Opt-Out Options}
\label{app:opt-out}

All the opt-out options selected directly on the two smart TVs are shown in Table \ref{tab:opt-out-options}.

\section{Amount of Bytes Towards ACR Domains}
\label{app:bytes}

Tables \ref{tab:bytes_acr}, \ref{tab:bytes_acr_P2}, \ref{tab:bytes_acr_US_P1} and \ref{tab:bytes_acr_US_P2} quantify the amount of data (kilobytes) exchanged with LG and Samsung ACR destinations across various scenarios. 

\begin{figure*}
	\centering
    \subfigure[LG]{
        \includegraphics[width=1\columnwidth]{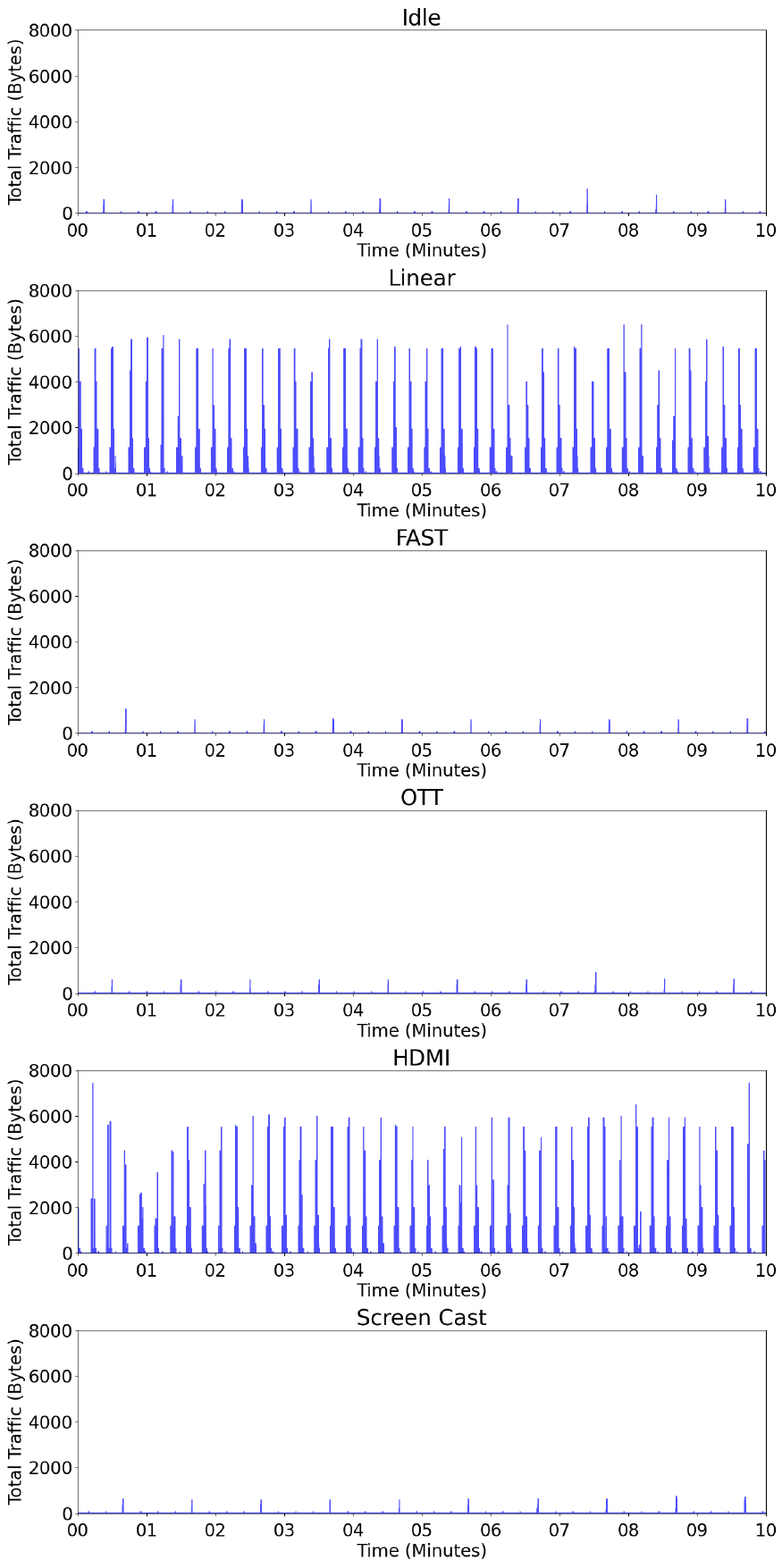}  
    }
	\centering
    \subfigure[Samsung]{
        \includegraphics[width=1\columnwidth]{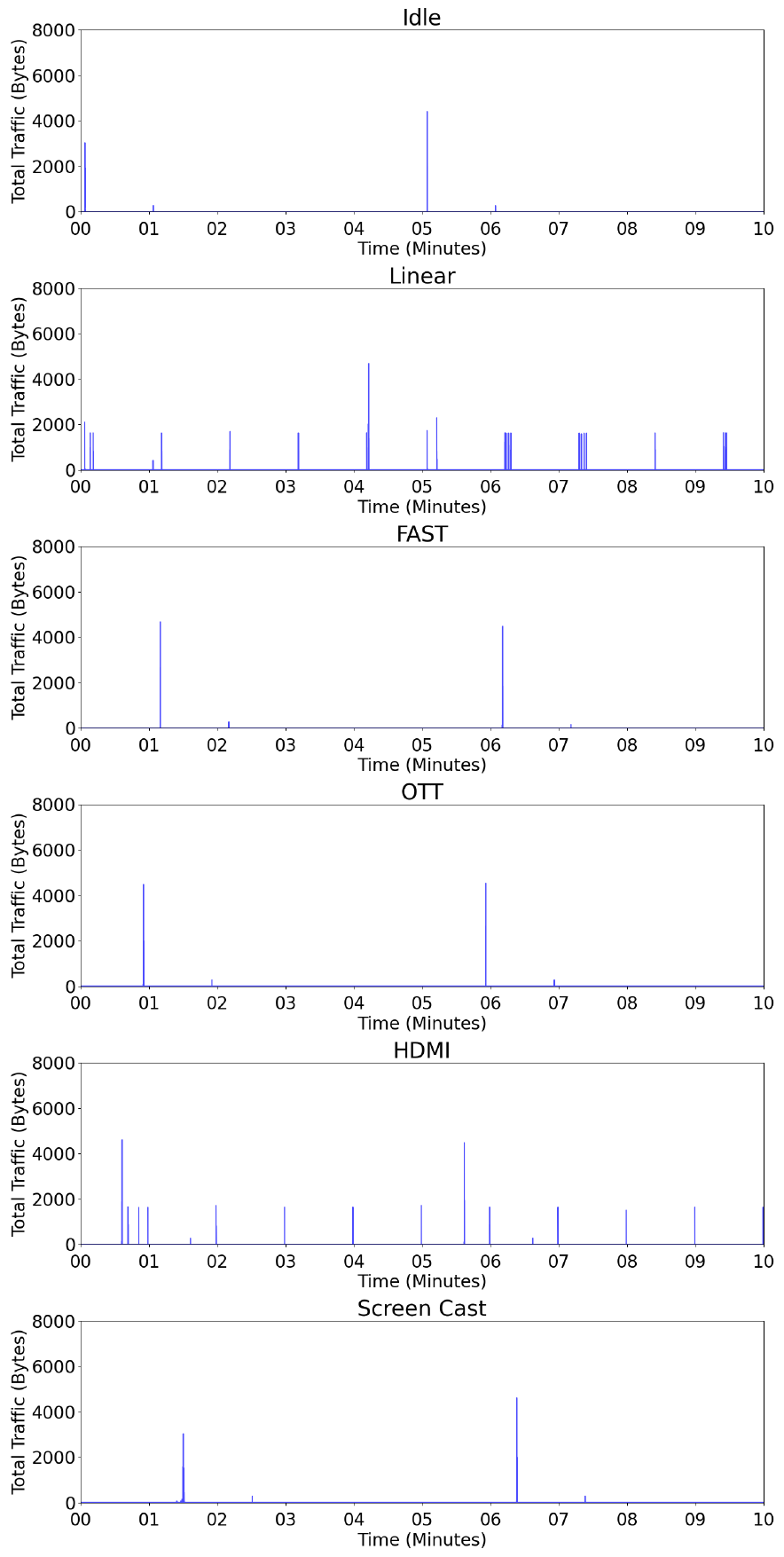} 
    }
  \caption{10 minutes of ACR traffic in different scenarios during LIn-OIn in UK.}
  \label{fig:UKP1-all}
\end{figure*}

\begin{figure*}
	\centering
    \subfigure[LG]{
        \includegraphics[width=1\columnwidth]{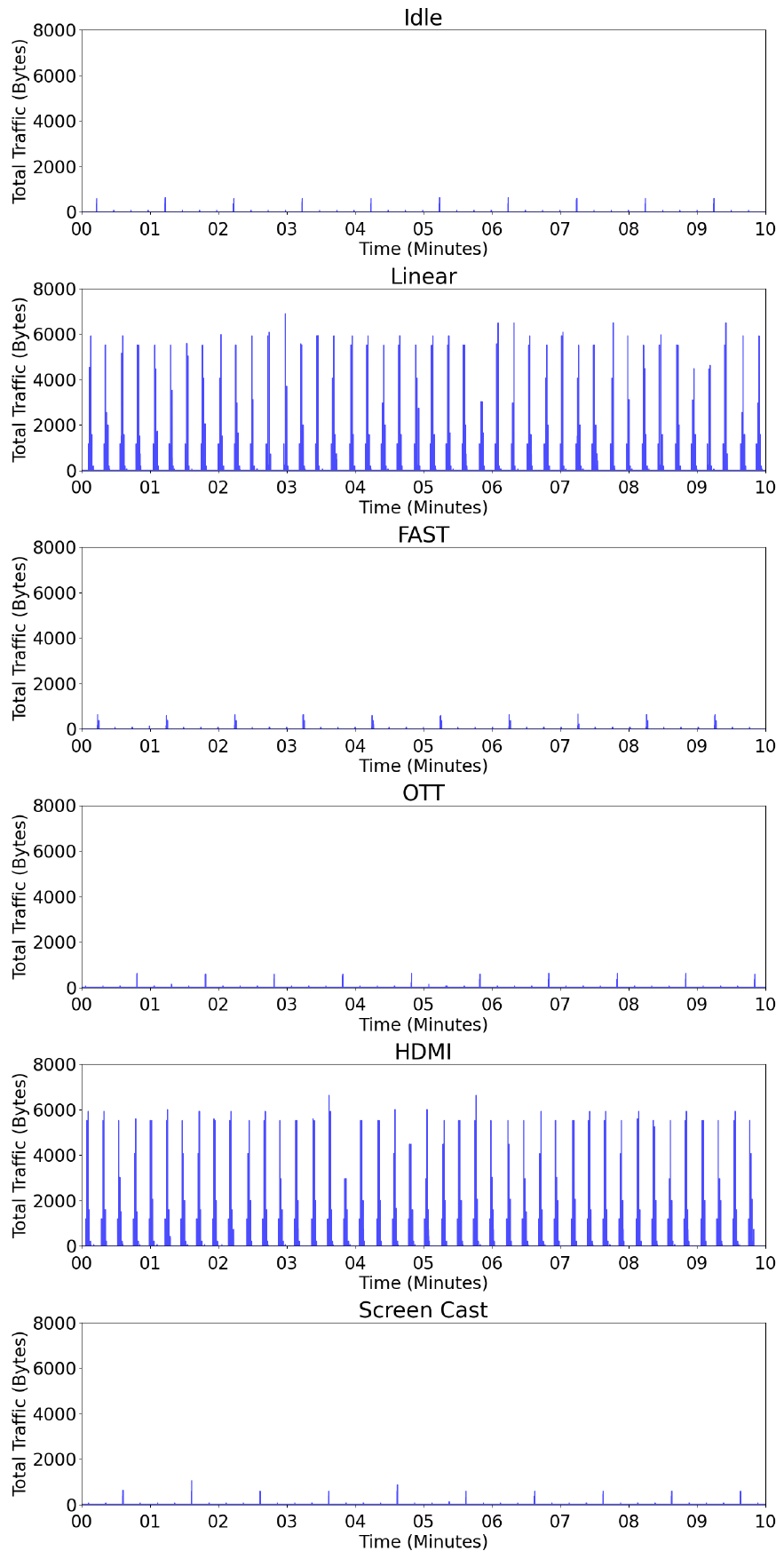}  
    }
	\centering
    \subfigure[Samsung]{
        \includegraphics[width=1\columnwidth]{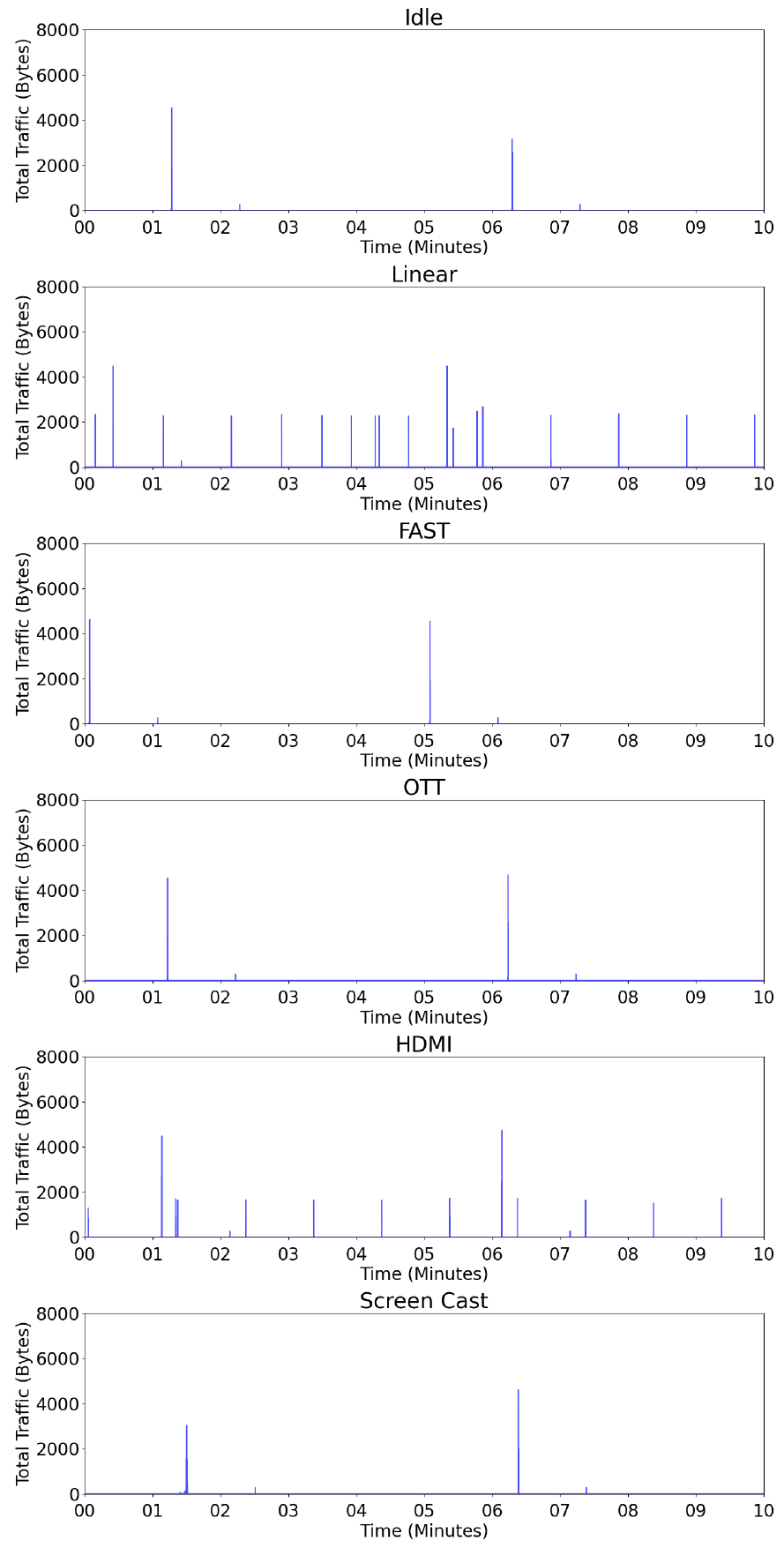} 
    }
  \caption{10 minutes of ACR traffic in different scenarios during LOut-OIn in UK.}
  \label{fig:UKP2-all}
\end{figure*}

\begin{figure*}
	\centering
    \subfigure[LG]{
        \includegraphics[width=1\columnwidth]{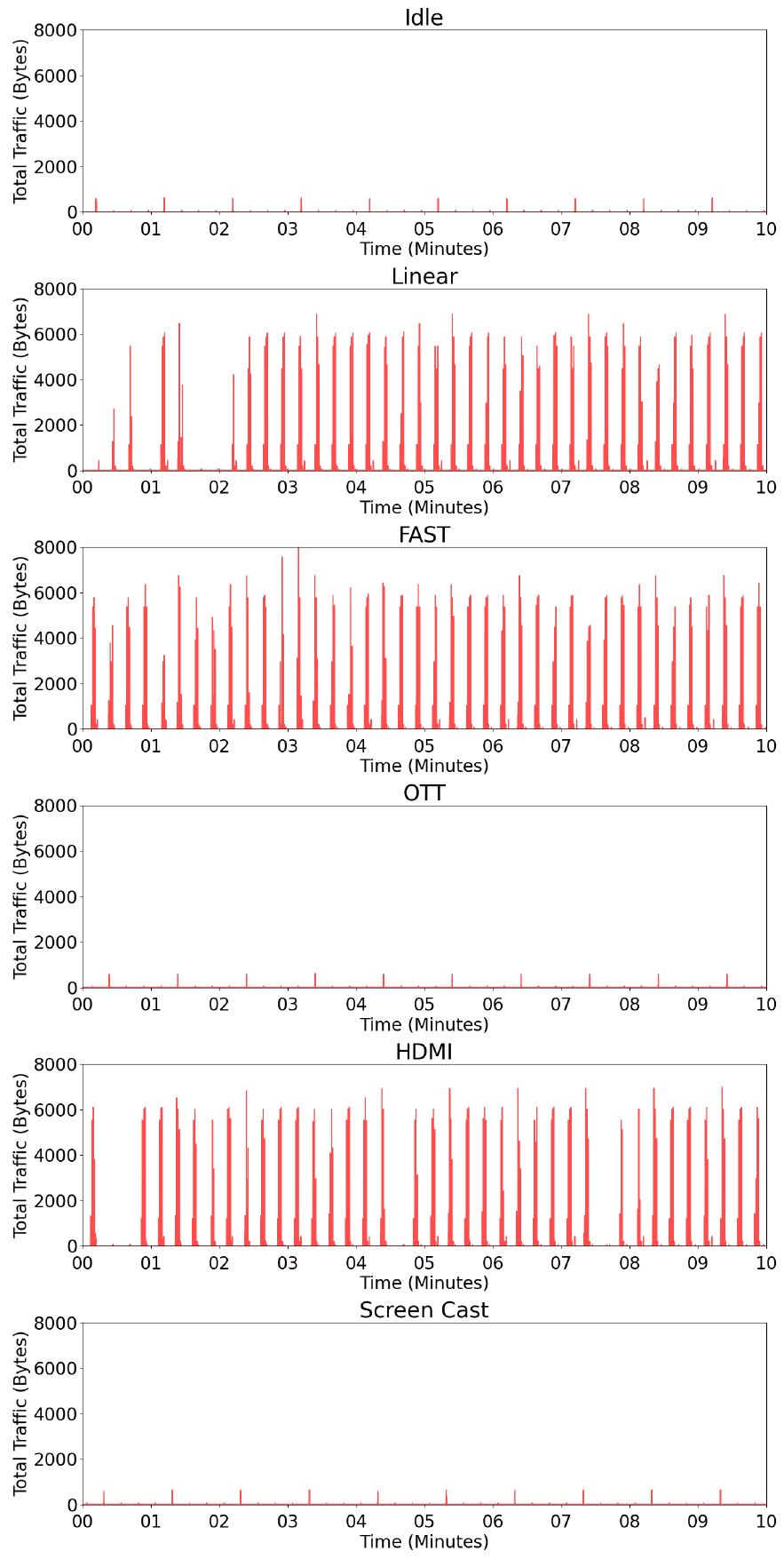}  
    }
	\centering
    \subfigure[Samsung]{
        \includegraphics[width=1\columnwidth]{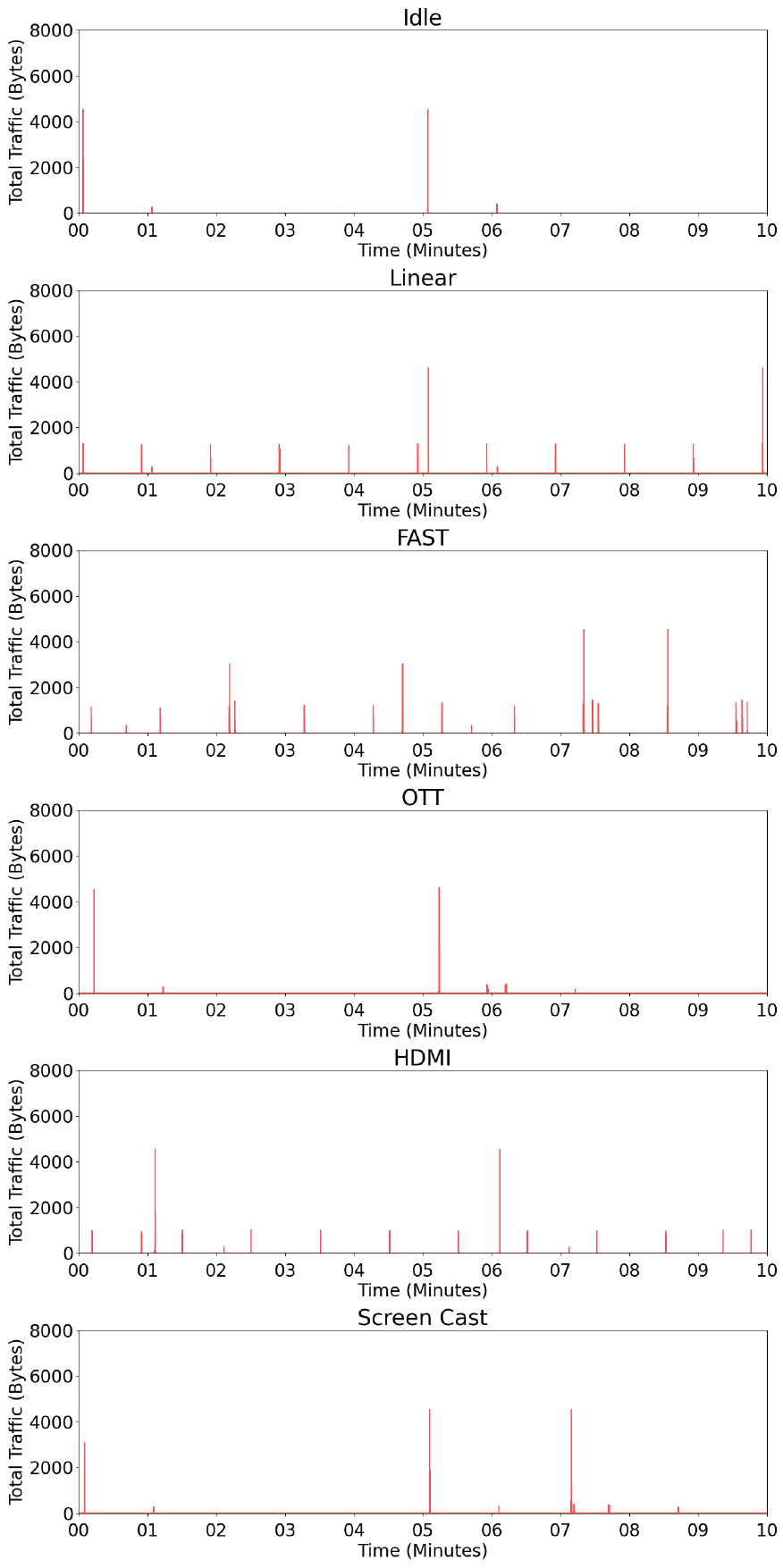} 
    }
  \caption{10 minutes of ACR traffic in different scenarios during LIn-OIn in US.}
  \label{fig:USP1-all}
\end{figure*}

\begin{figure*}
	\centering
    \subfigure[LG]{
        \includegraphics[width=1\columnwidth]{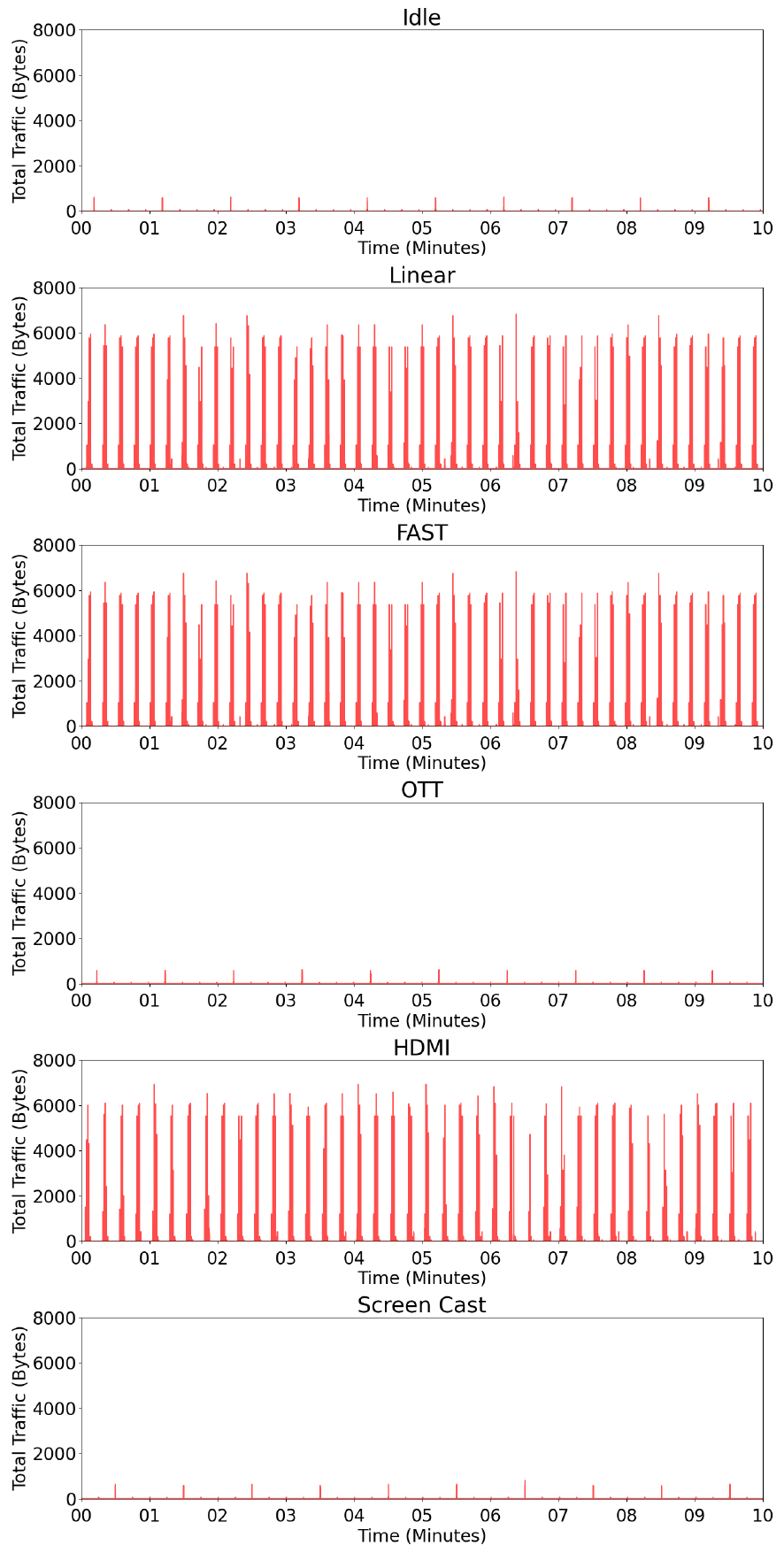}  
    }
	\centering
    \subfigure[Samsung]{
        \includegraphics[width=1\columnwidth]{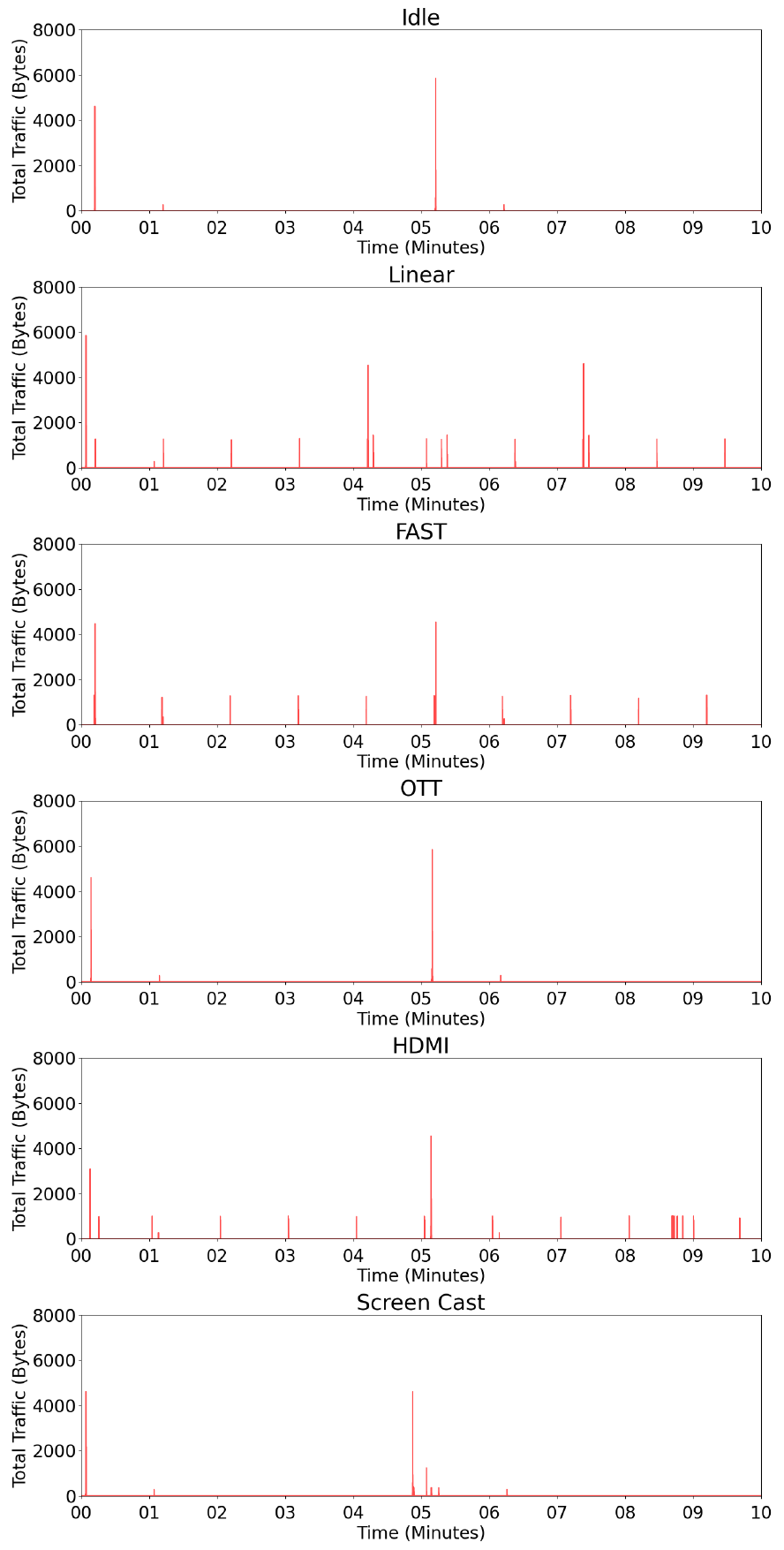} 
    }
  \caption{10 minutes of ACR traffic in different scenarios during LOut-OIn in US.}
  \label{fig:USP2-all}
\end{figure*}

Figures \ref{fig:UKP1-all}, \ref{fig:UKP2-all}, \ref{fig:USP1-all} and \ref{fig:USP2-all} show the details of ten minutes traffic for each presented scenarios, respectively for UK LIn-OIn, UK LOut-OIn, US LIn-OIn and US LOut-OIn.

\end{document}